*Article*

# Modeling of Breaching Due to Overtopping Flow and Waves Based on Coupled Flow and Sediment Transport

**Zhiguo He [1], Peng Hu [1], Liang Zhao [1], Gangfeng Wu [2,*] and Thomas Pähtz [1]**

[1] Ocean College, Zhejiang University, Hangzhou 310058, China;
E-Mails: hezhiguo@zju.edu.cn (Z.H.); penghu@zju.edu.cn (P.H.); liangz@zju.edu.cn (L.Z.); 0012136@zju.edu.cn (T.P.)

[2] School of Civil Engineering and Architecture, Ningbo Institute of Technology, Zhejiang University, Ningbo 31510, China

**\*** Author to whom correspondence should be addressed; E-Mail: 05szywgf@st.zju.edu.cn; Tel.: +86-571-8820-8912.

**Abstract:** Breaching of earthen or sandy dams/dunes by overtopping flow and waves is a complicated process with strong, unsteady flow, high sediment transport, and rapid bed changes in which the interactions between flow and morphology should not be ignored. This study presents a depth-averaged two-dimensional (2D) coupled flow and sediment transport model to investigate the flow and breaching processes with and without waves. Bed change and variable flow density are included in the flow continuity and momentum equations to consider the impacts of sediment transport. The model adopts the non-equilibrium approach for total-load sediment transport and specifies different repose angles to handle non-cohesive embankment slope avalanching. The equations are solved using an explicit finite volume method on a rectangular grid with the improved Godunov-type central upwind scheme and the nonnegative reconstruction of the water depth method to handle mixed-regime flows near the breach. The model has been tested against two sets of experimental data which show that it well simulates the flow characteristics, bed changes, and sediment transport. It is then applied to analyze flow and morphologic changes by overtopping flow with and without waves. The simulated bed change and breach cross-section shape show a significant difference if waves are considered. Erosion by flow without waves mainly occurs at the breach and is dominated by vertical erosion at the initial stage followed by the lateral erosion. With waves, the flow overtops the entire length of the dune to cause faster erosion along the entire length. Erosion mainly takes place at the upper layer at the initial stage and gradually accelerates as the height of the dune reduces and flow discharge increases, which indicates the simulated results with waves shall be further verified by physical experimental evidence.

## 1. Introduction

The failure of dams/levees often lead to catastrophic floods that cause loss of life and damage to property and agriculture. Overtopping flow has been one of the most common reasons for dam/levee failures, especially when heavy rainfall inland or storm surges along the coastline occur [1]. During such extreme events, wind-generated waves may run up the face of the dam/levee and enhance overtopping, which makes the erosion and breaching process more complicated [2,3]. Consequently, understanding the characteristics of flow, sediment transport, and the dam/levee breach process by overtopping flow and waves, and predicting discharge, breach width, depth, and sediment deposition have great significance for the decision-making in flood prevention and risk analysis.

In recent decades, the dam/levee breach problem has been extensively studied in mechanism, experiment, field, and modeling. A good review and summary can be found in the recent publication of the American Society of Civil Engineers/Environmental & Water Resources Institute (ASCE/EWRI) Task Committee on Dam/Levee Breaching [4]. However, the breaching process and its rapid morphologic changes due to overtopping flow and waves are still little understood due to observing and measuring difficulties in either lab experiments or field investigations. Even fewer results are available for the breach characteristics, such as breach width, depth, section shape, and bed changes, downstream at different time stages to examine the influence of waves during overtopping flows.

Two-dimensional (2D) numerical models solving the nonlinear shallow water equations (SWEs), the sediment transport, and bed change equations have been widely used to mathematically describe free surface flows and their associated morphologic changes over complex topography. Those models are often discretized by finite volume methods or finite difference methods and solved by Roe, Harten, Lax, and van Leer (HLL)/HLLC(C stands for Contact) [5], or TVD (total variation diminishing) schemes [6] in order to simulate the flood shock wave. However, simulations of earthen or sandy breaching processes due to overtopping flow and waves are much more difficult than those over fixed beds because highly unsteady flows, high concentrations of sediment, and rapid changes of bed can occur.

In recent years, a considerable number of studies proposed a coupled flow and sediment modeling concept that takes into account the impacts of bed change and variation of sediment concentration on the flow when simulating flows over movable beds [7–11]. These studies suggested that the flow continuity and momentum equations should include these impacts and the flow, sediment transport, and bed change should be calculated in a coupled approach. Meanwhile, non-equilibrium sediment transport approaches have been recommended to solve the cases of transient flows in which strong erosion and deposition occur [7,10–12], since the traditional assumption of local equilibrium capacity of sediment transport could be problematic. It might be inappropriate to use the sediment entrainment or transport capacity functions developed based on uniform flow conditions for embankment breach flow situations [7,10,13]. Furthermore, the bank erosion or collapse calculation should also be included in the modeling frame to deal with the instability of slopes [13].

To more accurately simulate the breaching and sediment transport processes and analyze their features due to overtopping flow and wave overwash, an improved cell-centered finite volume model based on the Godunov-type central upwind scheme [14] is developed to simulate shallow water flow and sediment transport over arbitrary topography. The model uses a non-equilibrium sediment transport approach and takes into account the effects of sediment concentration and bed change in the flow governing equations. The model solves the flow, sediment transport, and bed change equations in a coupled way based on the structured mesh.

The governing equations and the numerical methods, including the central upwind scheme and nonnegative water depth reconstruction, are presented in Sections 2 and 3, respectively. The model is then tested in Section 4 against measurements and afterwards applied to investigate the flow and erosion processes in the cases of overtopping flow with and without waves in Section 5.

## 2. Governing Equations of Flow and Sediment Transport in Breaching Process

During dam/levee breach, the sediment concentrations are high and beds change rapidly. Thus, the interactions between flow, sediment transport, and channel bed cannot be ignored. We take into account the bed changes and density variability of flow and sediment mixture in mass and momentum conservation equations based on the non-equilibrium concept [7,15], which assumes that sediment cannot reach new equilibrium states instantaneously due to the temporal and spatial lags between flow and sediment transport. This assumption is more realistic, especially in cases of strong erosion and deposition. Therefore, considering that the flow density is variable in the vertical direction, the depth-averaged 2D equations governing flow and total load sediment transport processes due to overtopping flow can be expressed as [7]

$$\frac{\partial(\rho h)}{\partial t} + \frac{\partial(\rho u h)}{\partial x} + \frac{\partial(\rho v h)}{\partial y} + \rho_b \frac{\partial z_b}{\partial t} = 0 \tag{1}$$

$$\frac{\partial}{\partial t}(\rho u h) + \frac{\partial}{\partial x}(\rho u^2 h) + \frac{\partial}{\partial y}(\rho u v h) + \rho g h \frac{\partial z_s}{\partial x} + \frac{1}{2} g h^2 \frac{\partial \rho}{\partial x} + \rho g \frac{n^2 m_b u U}{h^{1/3}} = 0 \tag{2}$$

$$\frac{\partial}{\partial t}(\rho v h) + \frac{\partial}{\partial x}(\rho u v h) + \frac{\partial}{\partial y}(\rho v^2 h) + \rho g h \frac{\partial z_s}{\partial y} + \frac{1}{2} g h^2 \frac{\partial \rho}{\partial y} + \rho g \frac{n^2 v U}{h^{1/3}} = 0 \tag{3}$$

$$\frac{\partial(h C_t)}{\partial t} + \frac{\partial(h u C_t)}{\partial x} + \frac{\partial(h v C_t)}{\partial y} = -\frac{1}{L_t}(U h C_t - m_b q_{t*}) \tag{4}$$

where $t$ is the time; $x$ and $y$ are the longitudinal and lateral coordinates; $h$ is the flow depth; $u$ and $v$ are the flow velocities in $x$ and $y$ directions, $U = \sqrt{u^2 + v^2}$; $\partial z_b/\partial t$ denotes the rate of change of $z_b$ (bed surface elevation above a reference datum); $z_s$ is the water level, $z_s = z_b + h$; $n$ is the Manning roughness coefficient; $g$ is the gravitational acceleration; $C_t$ is the actual volumetric total-load sediment concentration; $\rho$ is the density of the water and sediment mixture in the water column determined by $\rho = \rho_w(1 - C_t) + \rho_s C_t$; $\rho_b$ is the density of the water and sediment mixture in the bed surface layer determined by $\rho_b = \rho_w p'_m + \rho_s (1 - p'_m)$ with $p'_m$ being the porosity of the surface-layer bed material, $\rho_w$ and $\rho_s$ are the water and sediment densities, respectively; $m_b = \sqrt{1 + (\partial z_b / \partial x)^2 + (\partial z_b / \partial y)^2}$ (considers the lateral erosion along the side slope); $q_{t*}$ is the total-load sediment transport capacity, also called the equilibrium transport rate per unit channel width. $L_t$ is the non-equilibrium adaptation length or saturation length of sediment transport [15–17]. The total-load sediment transport capacity and the adaptation length are determined by the same method proposed by Wu [15].

Using the relations $\rho = \rho_w (1 - C_t) + \rho_s C_t$, $z_s = z_b + h$, and Equation (4), the flow density on the left-hand sides of Equations (1)–(3) can be separated from the variables. Thus, Equations (1)–(4) can be rewritten in the following compact form as [15]

$$\frac{\partial \mathbf{\Phi}}{\partial t} + \frac{\partial \mathbf{F}(\mathbf{\Phi})}{\partial x} + \frac{\partial \mathbf{G}(\mathbf{\Phi})}{\partial y} = \mathbf{S}(\mathbf{\Phi}) \tag{5}$$

where $\mathbf{\Phi}$, $\mathbf{F}(\mathbf{\Phi})$, and $\mathbf{G}(\mathbf{\Phi})$ represent the vectors of unknown variables, fluxes, and source terms, respectively:

$$\mathbf{\Phi} = \begin{bmatrix} h \\ uh \\ vh \\ hC_t \end{bmatrix}, \quad \mathbf{F}(\mathbf{\Phi}) = \begin{bmatrix} uh \\ u^2h + \frac{1}{2}gh^2 \\ uvh \\ huC_t \end{bmatrix}, \quad \mathbf{G}(\mathbf{\Phi}) = \begin{bmatrix} vh \\ uvh \\ v^2h + \frac{1}{2}gh^2 \\ hvC_t \end{bmatrix} \tag{6}$$

$$\mathbf{S} = \begin{bmatrix} -\frac{1}{1-p'_m} \frac{UhC_t - m_b q_{t*}}{L_t} \\ -gh\frac{\partial z_b}{\partial x} - \frac{1}{2}gh^2 \frac{1}{\rho}\frac{\partial \rho}{\partial x} - g\frac{n^2 m_b u U}{h^{1/3}} + u\frac{\rho_s - \rho_w}{\rho}\left(1 - \frac{C_t}{1-p'_m}\right)\frac{UhC_t - m_b q_{t*}}{L_t} \\ -gh\frac{\partial z_b}{\partial y} - \frac{1}{2}gh^2 \frac{1}{\rho}\frac{\partial \rho}{\partial y} - g\frac{n^2 m_b v U}{h^{1/3}} + v\frac{\rho_s - \rho_w}{\rho}\left(1 - \frac{C_t}{1-p'_m}\right)\frac{UhC_t - m_b q_{t*}}{L_t} \\ -\frac{1}{L_t}(UhC_t - m_b q_{t*}) \end{bmatrix} \tag{7}$$

The bed deformation is determined with

$$(1 - p'_m)\frac{\partial z_b}{\partial t} = \frac{1}{L_t}(UhC_t - m_b q_{t*}) \tag{8}$$

Moreover, the sediment transport capacity is determined using the formula of Wu *et al.* [18] and the effect of gravity on sediment transport over steep slopes is also considered [19]. The settling velocity $\omega_s$ used in Wu *et al.*'s [18] formula is determined by the Richardson [20] method: $\omega_s = \omega_{s0}(1 - C_t)^m$, where m ≈ 4.0 and $\omega_{s0}$ is the settling velocity of particles in still and clear water.

## 3. Numerical Methods

*3.1. Model Discretization*

The mathematical model is discretized using a finite volume method based on a rectangular mesh, as shown in Figure 1, in which a non-staggered (collocated) grid system is used. The primary variables $h$, $u$, $v$, and $C_t$, and the bed elevation are defined at cell centers, representing the average values over each cell. The fluxes are calculated at the cell faces.

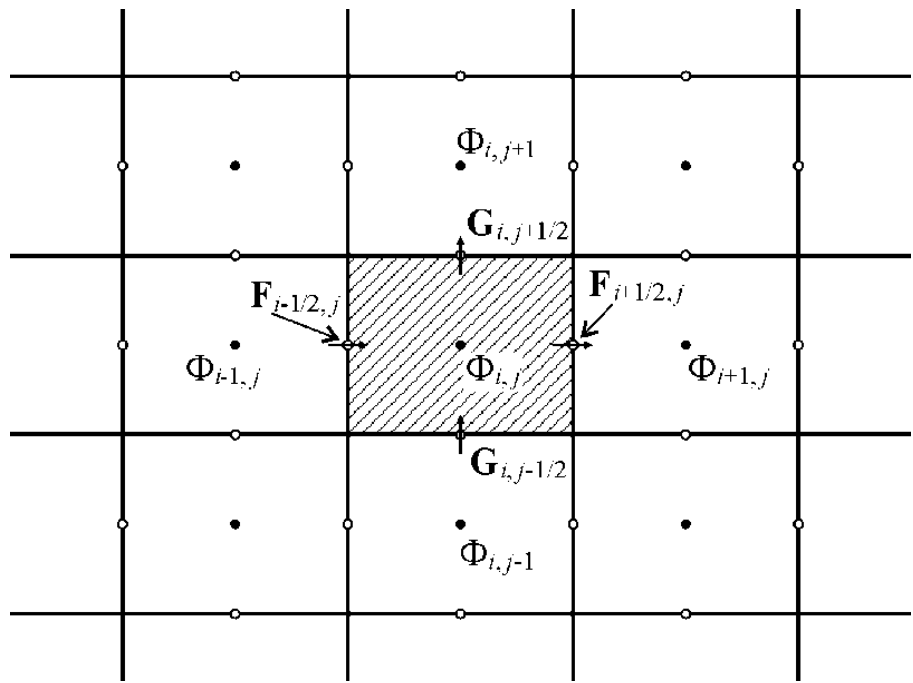

**Figure 1.** 2D finite-volume mesh.

Integrating Equation (5) over the $(i, j)$ control volume and applying the Euler scheme, Equation (5) can be discretized using Green's theorem as follows:

$$\Phi_{i,j}^{n+1} = \Phi_{i,j}^{n} - \frac{\Delta t}{\Delta x_{i,j}} \left( \mathbf{F}_{i+1/2,j}^{n} - \mathbf{F}_{i-1/2,j}^{n} \right) - \frac{\Delta t}{\Delta y_{i,j}} \left( \mathbf{G}_{i,j+1/2}^{n} - \mathbf{G}_{i,j-1/2}^{n} \right) + \Delta t \, \mathbf{S}_{i,j} \qquad (9)$$

where $\Delta t$ is the time step length, $\Delta x_{i,j}$ and $\Delta y_{i,j}$ are the cell lengths in $x$ and $y$ directions, $\mathbf{F}_{i+1/2,j}^{n}$ is the flux at face $(i+1/2, j)$, and $\mathbf{G}_{i,j+1/2}^{n}$ is the flux at face $(i, j+1/2)$. $\mathbf{S}_{i,j}$ represents the source terms evaluated at the cell center.

The determination of the fluxes at the cell faces is one of most important procedures to solve Equation (9). Thus, this paper develops an improved mass-balanced 2D shallow water and sediment transport model based on the Godunov-type central upwind scheme [14] and the nonnegative reconstruction of the water depth method [21].

*3.2. Nonnegative Reconstruction of Riemann State*

A correct reconstruction of the Riemann states [6] at the interfaces is the key to determine the interface fluxes accurately. Here the nonnegative reconstruction of the water depth method [21] is adopted to calculate the Riemann states, which have been proved to be accurate to deal with wetting and drying. The Riemann states, state variables, and fluxes defined at the interface can be obtained from the available flow information at the cell center by using a slope-limited linear reconstruction, e.g., left-hand side values at the interface $(i + 1/2, j)$ are computed by

$$\overline{\eta}^L_{i+1/2,j} = \eta_{i,j} + \frac{\Delta x}{2}(\eta_x)_{i,j}, \quad \overline{h}^L_{i+1/2,j} = h_{i,j} + \frac{\Delta x}{2}(h_x)_{i,j}, \quad (\overline{z}_b)^L_{i+1/2,j} = \overline{\eta}^L_{i+1/2,j} - \overline{h}^L_{i+1/2,j} \tag{10a}$$

$$(\overline{hu})^L_{i+1/2,j} = (hu)_{i,j} + \frac{\Delta x}{2}((hu)_x)_{i,j}, \quad (\overline{hv})^L_{i+1/2,j} = (hv)_{i,j} + \frac{\Delta x}{2}((hv)_x)_{i,j} \tag{10b}$$

where $(\eta_x)_{i,j}$, $(h_x)_{i,j}$, $((hu)_x)_{i,j}$, and $((hv)_x)_{i,j}$ represent the numerical derivatives of the flow variable at cell $(i, j)$. In order to preserve the stability of the model, a one-parameter family of generalized min mod limiters is used to calculate the numerical derivatives. For example,

$$(\eta_x)_{i,j} = \text{minmod}(\theta \frac{\eta_{i,j} - \eta_{i-1,j}}{\Delta x}, \frac{\eta_{i+1,j} - \eta_{i-1,j}}{2\Delta x}, \theta \frac{\eta_{i+1,j} - \eta_{i,j}}{\Delta x}), \theta \in [1, 2] \tag{11}$$

where θ is the parameter to control numerical viscosity. In general, larger values of θ make the numerical result less dissipative and more oscillatory [6]. θ = 1.3, suggested by Kurganov and Petrova [22], is used in this study. The min mod function is defined as

$$\text{minmod}(z_1, z_2, \ldots, z_n) = \begin{cases} \max z_i, & z_i < 0 \ \forall i, \\ \min z_i, & z_i > 0 \ \forall i, \\ 0, & \text{otherwize}. \end{cases} \tag{12}$$

The value of $(h_x)_{i,j}$, $((hu)_x)_{i,j}$, and $((hv)_x)_{i,j}$ can be obtained in a similar way. Similarly, the right-hand side values at the interface $(i + 1/2, j)$ can be calculated by

$$\overline{\eta}^R_{i+1/2,j} = \eta_{i+1,j} - \frac{\Delta x}{2}(\eta_x)_{i+1,j}, \quad \overline{h}^R_{i+1/2,j} = h_{i+1,j} - \frac{\Delta x}{2}(h_x)_{i+1,j}, \quad (\overline{z}_b)^R_{i+1/2,j} = \overline{\eta}^R_{i+1/2,j} - \overline{h}^R_{i+1/2,j} \tag{13a}$$

$$(\overline{hu})^R_{i+1/2,j} = (hu)_{i+1,j} - \frac{\Delta x}{2}((hu)_x)_{i+1,j}, \quad (\overline{hv})^R_{i+1/2,j} = (hv)_{i+1,j} - \frac{\Delta x}{2}((hv)_x)_{i+1,j} \tag{13b}$$

To obtain the final Riemann states, a single value of the bed elevation suggested by Audusse *et al.* [23] should be defined as

$$(z_b)_{i+1/2,j} = \max((\overline{z}_b)^L_{i+1/2,j}, (\overline{z}_b)^R_{i+1/2,j}) \tag{14}$$

Then the face values of water depth are redefined as

$$h^L_{i+1/2,j} = \max(0, \overline{\eta}^L_{i+1/2,j} - (z_b)_{i+1/2,j}), \quad h^R_{i+1/2,j} = \max(0, \overline{\eta}^R_{i+1/2,j} - (z_b)_{i+1/2,j}) \tag{15}$$

Obviously, this method ensures the reconstructed water depth to be nonnegative. However, the reconstructed water depth may be very small or zero and thus lead to large values of the flow velocity, which may cause the model to be unstable. In order to prevent this problem, the regularization technique suggested by Kurganov and Petrova [22] is used to calculate the corresponding velocity components at the interface:

$$u^L_{i+1/2,j} = \frac{\sqrt{2}h^L_{i+1/2,j} \cdot \overline{(hu)}^L_{i+1/2,j}}{\sqrt{(h^L_{i+1/2,j})^4 + \max((h^L_{i+1/2,j})^4, \varepsilon)}}, \quad v^L_{i+1/2,j} = \frac{\sqrt{2}h^L_{i+1/2,j} \cdot \overline{(hv)}^L_{i+1/2,j}}{\sqrt{(h^L_{i+1/2,j})^4 + \max((h^L_{i+1/2,j})^4, \varepsilon)}} \quad (16a)$$

$$u^R_{i+1/2,j} = \frac{\sqrt{2}h^R_{i+1/2,j} \cdot \overline{(hu)}^R_{i+1/2,j}}{\sqrt{(h^R_{i+1/2,j})^4 + \max((h^R_{i+1/2,j})^4, \varepsilon)}}, \quad v^R_{i+1/2,j} = \frac{\sqrt{2}h^R_{i+1/2,j} \cdot \overline{(hv)}^R_{i+1/2,j}}{\sqrt{(h^R_{i+1/2,j})^4 + \max((h^R_{i+1/2,j})^4, \varepsilon)}} \quad (16b)$$

where ε is an empirically pre-defined positive number (ε = 10$^{-6}$ m in all our simulations).

Hence, the Riemann states of the other flow variables can be recalculated by

$$\eta^L_{i+1/2,j} = (z_b)_{i+1/2,j} + h^L_{i+1/2,j}, \quad (hu)^L_{i+1/2,j} = h^L_{i+1/2,j} u^L_{i+1/2,j}, \quad (hv)^L_{i+1/2,j} = h^L_{i+1/2,j} v^L_{i+1/2,j} \quad (17a)$$

$$\eta^R_{i+1/2,j} = (z_b)_{i+1/2,j} + h^R_{i+1/2,j}, \quad (hu)^R_{i+1/2,j} = h^R_{i+1/2,j} u^R_{i+1/2,j}, \quad (hv)^R_{i+1/2,j} = h^R_{i+1/2,j} v^R_{i+1/2,j} \quad (17b)$$

The procedure above obviously does not affect the well-mass-balanced property of the numerical scheme for wet-bed applications. However, the local bed modifications are needed to get the well-balanced solution for a dry-bed application. As shown in Figure 2, the wet cell ($i, j$) shares the common edge ($i + 1/2, j$) with the dry cell ($i + 1, j$), and the bed elevation of the dry cell is higher than the water surface elevation at cell ($i, j$). The problem of stationary water at the steady state is considered through $u \equiv 0$, $v \equiv 0$, and $\eta \equiv$ const at the wet area. After the aforementioned reconstruction at the cell interface ($i + 1/2, j$), the components of the Riemann states are given by

$$(z_b)_{i+1/2,j} = (\overline{z}_b)^R_{i+1/2,j}, \quad h^L_{i+1/2,j} = h^R_{i+1/2,j} = 0, \quad \eta^L_{i+1/2,j} = \eta^R_{i+1/2,j} = (z_b)_{i+1/2,j} \quad (18)$$

Therefore, the fluxes at cell interface ($i + 1/2, j$) are calculated using the bed elevation $(z_b)_{i+1/2,j}$ rather than the actual water surface elevation $\eta$ in cell ($i, j$). However, fluxes at the cell interface ($i - 1/2, j$) are estimated using the actual water surface elevation $\eta$ in cell ($i, j$), which produces a net flux in cell ($i, j$) and violates the well-balanced property of the scheme. The local bed modification is thus implemented to overcome this difficulty following Liang [21]. As show in Figure 2, the difference between the actual and fake water at interface ($i + 1/2, j$) is calculated by

$$\Delta z = \max\{0, (z_b)_{i+1/2,j} - \eta^L_{i+1/2,j}, (z_b)_{i+1/2,j} - \eta^R_{i+1/2,j}\} \quad (19)$$

The bed elevation and water surface are locally modified by subtracting $\Delta z$ from the original reconstructed value:

$$(z_b)_{i+1/2,j} \leftarrow (z_b)_{i+1/2,j} - \Delta z, \quad (\eta)^L_{i+1/2,j} \leftarrow (\eta)^L_{i+1/2,j} - \Delta z, \quad (\eta)^R_{i+1/2,j} \leftarrow (\eta)^R_{i+1/2,j} - \Delta z \quad (20)$$

Obviously, $(z_b)_{i+1/2,j} = (\eta)^L_{i+1/2,j} = (\eta)^R_{i+1/2,j} = \eta$ after the local bed modification and no net fluxes are produced in cell ($i, j$), which ensures the scheme to be well-balanced for the dry-bed application.

*3.3. Central Upwind Scheme*

Once the left and right variables are estimated using the aforementioned reconstruction method, the interface flux can be calculated using the Godunov-type central upwind scheme as follows [14]:

$$\mathbf{f}_{i+1/2,j} = \frac{a^+_{i+1/2,j}\mathbf{f}(\mathbf{q}^L_{i+1/2,j},(z_b)_{i+1/2,j}) - a^-_{i+1/2,j}\mathbf{f}(\mathbf{q}^R_{i+1/2,j},(z_b)_{i+1/2,j})}{a^+_{i+1/2,j} - a^-_{i+1/2,j}}$$
$$+ \frac{a^+_{i+1/2,j}a^-_{i+1/2,j}}{a^+_{i+1/2,j} - a^-_{i+1/2,j}}(\mathbf{q}^R_{i+1/2,j} - \mathbf{q}^L_{i+1/2,j}) \quad (21)$$

$$\mathbf{g}_{i,j+1/2} = \frac{b^+_{i,j+1/2}\mathbf{f}(\mathbf{q}^L_{i,j+1/2},(z_b)_{i,j+1/2}) - b^-_{i,j+1/2}\mathbf{f}(\mathbf{q}^R_{i,j+1/2},(z_b)_{i,j+1/2})}{b^+_{i,j+1/2} - b^-_{i,j+1/2}}$$
$$+ \frac{b^+_{i,j+1/2}b^-_{i,j+1/2}}{b^+_{i,j+1/2} - b^-_{i,j+1/2}}(\mathbf{q}^R_{i,j+1/2} - \mathbf{q}^L_{i,j+1/2}) \quad (22)$$

where $\mathbf{q}^L_{i+1/2,j}$, $\mathbf{q}^R_{i+1/2,j}$ are the reconstructed Riemann states at the left- and right-hand side of interface ($i + 1/2, j$), respectively; $\mathbf{q}^L_{i,j+1/2}$, $\mathbf{q}^R_{i,j+1/2}$ represent the reconstructed Riemann states at the left- and right-hand side of interface ($i, j + 1/2$), respectively; $a^\pm_{i+1/2,j}$, $b^\pm_{i,j+1/2}$ are the one-sided local wave speeds in *x* and *y* directions, respectively, and can be calculated as follows:

$$a^+_{i+1/2,j} = \max\{u^R_{i+1/2,j} + \sqrt{gh^R_{i+1/2,j}}, u^L_{i+1/2,j} + \sqrt{gh^L_{i+1/2,j}}, 0\} \quad (23)$$

$$a^-_{i+1/2,j} = \min\{u^R_{i+1/2,j} - \sqrt{gh^R_{i+1/2,j}}, u^L_{i+1/2,j} - \sqrt{gh^L_{i+1/2,j}}, 0\} \quad (24)$$

$$b^+_{i,j+1/2} = \max\{v^R_{i,j+1/2} + \sqrt{gh^R_{i,j+1/2}}, v^L_{i,j+1/2} + \sqrt{gh^L_{i,j+1/2}}, 0\} \quad (25)$$

$$b^-_{i,j+1/2} = \min\{v^R_{i,j+1/2} - \sqrt{gh^R_{i,j+1/2}}, v^L_{i,j+1/2} - \sqrt{gh^L_{i,j+1/2}}, 0\} \quad (26)$$

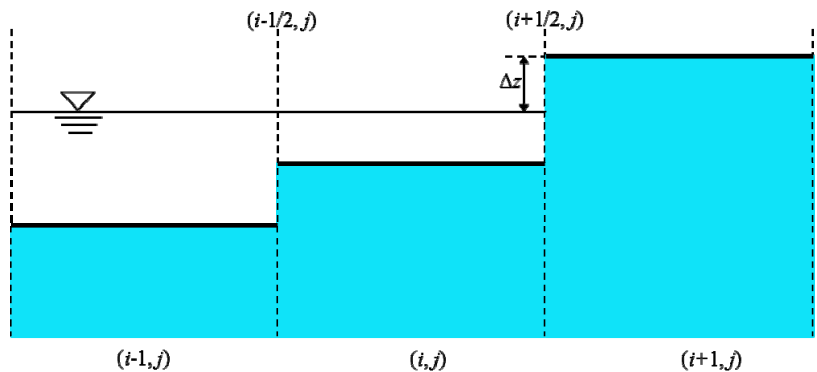

**Figure 2.** Local bed modification for dry-bed application.

*3.4. Treatments of the Source Terms*

The source terms, as shown in Equation (7), can be split into the bed slope terms and friction terms. It is important to discretize the bed slope terms appropriately to ensure the scheme to be well-balanced. Hence, the bed slope terms are approximated as follows [7]:

$$g(\eta - z_b)\frac{\partial z_b}{\partial x} = g\frac{(z_b)_{i+1/2,j} - (z_b)_{i-1/2,j}}{\Delta x} \cdot \frac{h^L_{i+1/2,j} + h^R_{i-1/2,j}}{2} \quad (27)$$

$$g(\eta - z_b)\frac{\partial z_b}{\partial y} = g\frac{(z_b)_{i,j+1/2} - (z_b)_{i,j-1/2}}{\Delta y} \cdot \frac{h^L_{i,j+1/2} + h^R_{i,j-1/2}}{2} \tag{28}$$

Herein the values of bed elevations are obtained after considering the local bed modification in Equation (19).

In general, a very small water depth may cause numerical instability when a simple, explicit discretization of the friction terms is used. To overcome this problem, implicit treatments [24] and semi-implicit treatments [25] are widely used. In this study, the friction terms are discretized using the semi-implicit treatment [7] given by:

$$S_{fx} = \frac{gu\sqrt{u^2+v^2}n^2}{h^{1/3}} = (\frac{g\sqrt{u^2+v^2}n^2}{h^{4/3}})^k (hu)^{k+1} \tag{29}$$

$$S_{fy} = \frac{gv\sqrt{u^2+v^2}n^2}{h^{1/3}} = (\frac{g\sqrt{u^2+v^2}n^2}{h^{4/3}})^k (hv)^{k+1} \tag{30}$$

where the superscript *k* represents the *k*th time level. Substituting Equations (29) and (30) into the time marching Equation (7), the values of the flow variables at the end of the time step can be obtained.

*3.5. Calculation of Sediment Transport*

The total-load sediment transport Equation (4) is solved using Equation (9), while the sediment flux at the cell faces is determined using the upwind scheme (e.g., flux at interface (*i* + 1/2, *j*)):

$$(huC_t)_{i+1/2,j} = (hu)^+_{i+1/2,j}(C_t)_{i,j} + (hu)^-_{i+1/2,j}(C_t)_{i+1,j} \tag{31}$$

where $(hu)^{\pm}_{i+1/2,j}$ is estimated by

$$(hu)^+_{i+1/2,j} = \max\{(hu)_{i+1/2,j}, 0.0\}, (hu)^-_{i+1/2,j} = \min\{(hu)_{i+1/2,j}, 0.0\} \tag{32}$$

The bed deformation in Equation (8) is discretized as

$$(z_b)^{o+1} = (z_b)^o + \frac{\Delta t}{L_t(1-p_m)}(hC_t\sqrt{u^2+v^2} - q_t) \tag{33}$$

$$\Delta z_b = \frac{\Delta t}{1-p'_m}\left(\frac{UhC_t - m_b q_{t*}}{L_t}\right)^n \tag{34}$$

In order to avoid instability of sediment transport, flow velocities at the cell centers are evaluated from the intercell fluxes as presented by Wu *et al.* [7] to determine the bed shear stress and sediment transport capacity $q_{t*}$.

The non-cohesive slope avalanching algorithm presented by Wu [15] is modified by specifying different angles of repose for the submerged materials and the dry materials above the water surface. The submerged repose angle is about 33 degrees as normally used, while the dry repose angle is about 79 degrees, as suggested by Wu [15], for the test cases presented in Section 4.

*3.6. Stability Criterion and Boundary Conditions*

The current numerical scheme is explicit, and its stability is governed by the Courant-Friedrichs-Lewy (CFL) criterion and additional conditions for sediment transport and bed change computations, as described in the following. In order to preserve the positivity of water depth at each time step, the Courant number has to be kept less than 0.25 [26], reading

$$N_{CFL} = \max\{\frac{a\Delta t}{\Delta x}, \frac{b\Delta t}{\Delta y}\} \leq 0.25 \qquad (35)$$

where $N_{CFL}$ represents the Courant number, and *a* and *b* are given by:

$$a = \max_{i,j}\{\max\{a^+_{i+1/2,j}, -a^-_{i+1/2,j}\}\}, \quad b = \max_{i,j}\{\max\{b^+_{i,j+1/2}, -b^-_{i,j+1/2}\}\} \qquad (36)$$

Moreover, the bed change at each time step should be less than about 10% of the local flow depth [7]. Thus, the CFL restriction for the current model is given as

$$\Delta t \leq \min[\,0.25\min\{\frac{\Delta x}{a}, \frac{\Delta y}{b}\}, \frac{0.1hL(1-p_m)}{\sqrt{u^2+v^2}hC_t - q_t}\,] \qquad (37)$$

Generally, two kinds of boundaries, open boundary and wall boundary conditions, are used in the numerical model. For the open boundary conditions, flow variables in the ghost cells are usually determined by solving a boundary Riemann problem, which is dependent on the local flow regime [27,28]. For wall boundary conditions, the velocity normal to the boundary and the water surface gradient are both set to zero at the boundary. It was shown that this procedure is second-order accurate [29].

**4. Model Verification**

*4.1. Case 1: Tsunami Run-Up onto a Complex Three-Dimensional (3D) Beach*

This laboratory experiment was suggested as a benchmark test for numerical models in The Third International Workshop on Long Wave Run-up Models in 2004. It was performed in a 1:400 scale wave tank, which approximated the coastline topography near Monai in Japan. This area was inundated by the 1993 Okushiri tsunami and a maximum runup of 31.7 m was observed during this event. As shown in Figure 3, the tank was 5.5 m long and 3.4 m wide. The incoming wave was generated by the movements of wave paddles and three gauges, Ch5, Ch7, and Ch9 (see Figure 3), were set up to record the time history of water elevation.

For this case, the computational domain is approximated made of 392 × 243 grids with a uniform grid size of 0.014 m. At the beginning, the tank is filled with still water of 0.135 m depth. The left boundary is the wave inflow boundary, in which the time history of the measured surface elevation was specified. The other three side boundaries are set to be closed. The total simulation time is 22.5 s and the Manning coefficient is *n* = 0.0025, which was suggested by [30]. The computed surface elevation at different times is plotted in Figure 4, while Figure 5 shows the computed water elevation at three gauges compared with the observation data. In general, the movement of the wave is well predicted by the present model. Some discrepancies are found between numerical and experimental results due to the fact that the three-dimensionality of the flow cannot be exactly captured by a depth-averaged 2D model. Nevertheless, the lead wave height (with relative errors less than 2.3%) and arrival times (with relative errors less than 0.5%) are predicted with a good accuracy at these gauges.

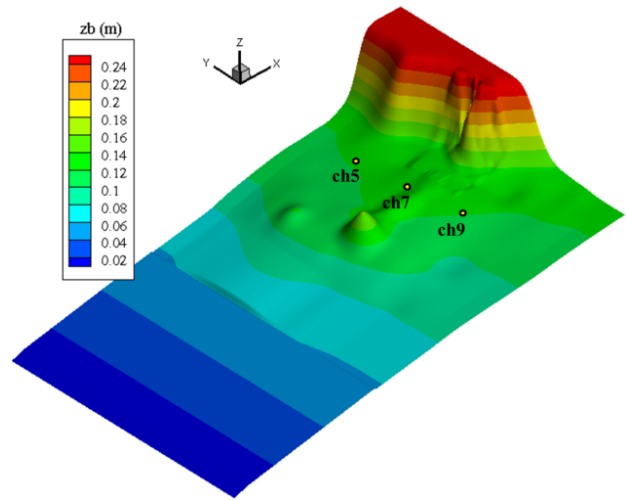

**Figure 3.** Three-dimensional view of bed elevation and the locations of the three gauges. Bed elevation is presented in the colored contours and dots represent three recoded gauges. Red color represents high land and blue represents sea side.

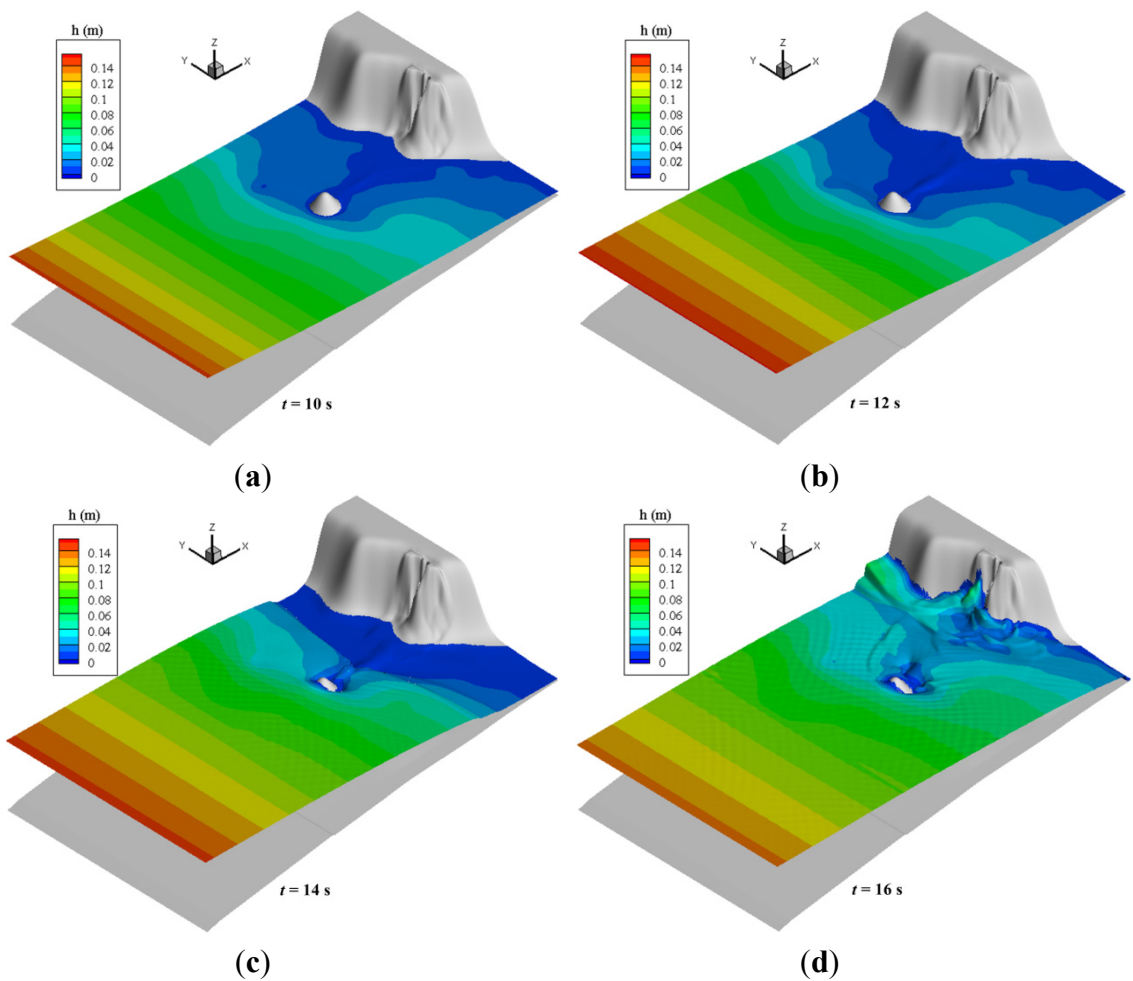

(a) $t = 10$ s   (b) $t = 12$ s

(c) $t = 14$ s   (d) $t = 16$ s

**Figure 4.** *Cont.*

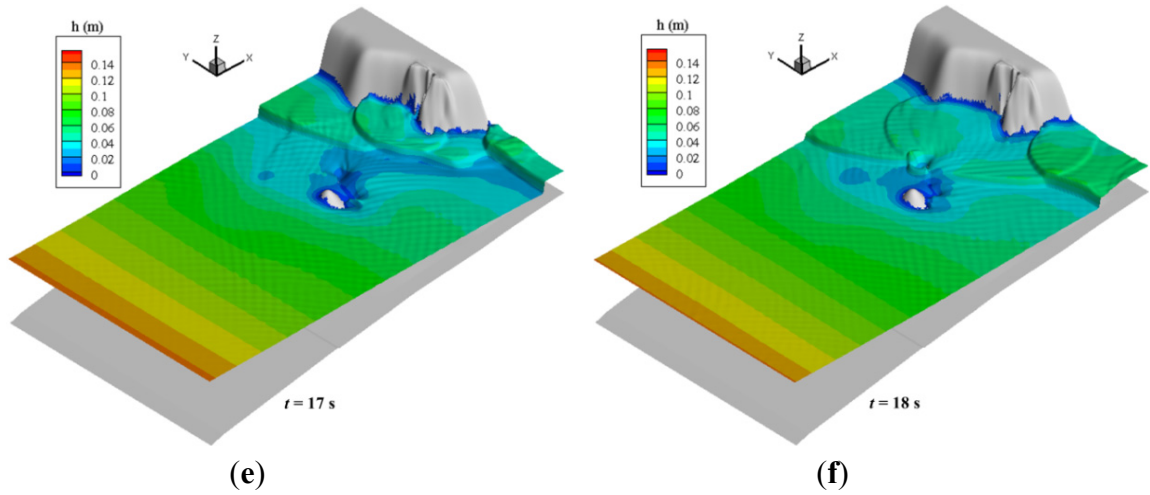

(e) (f)

**Figure 4.** The simulated inundation map at different times in a 3D view. Numerical results of water depth are presented in the colored contours, while grey color represents bed elevation. (**a**–**f**) represent water depth at the simulation times of 10 s, 12 s, 14, 16, 17 s, and 18 s, respectively.

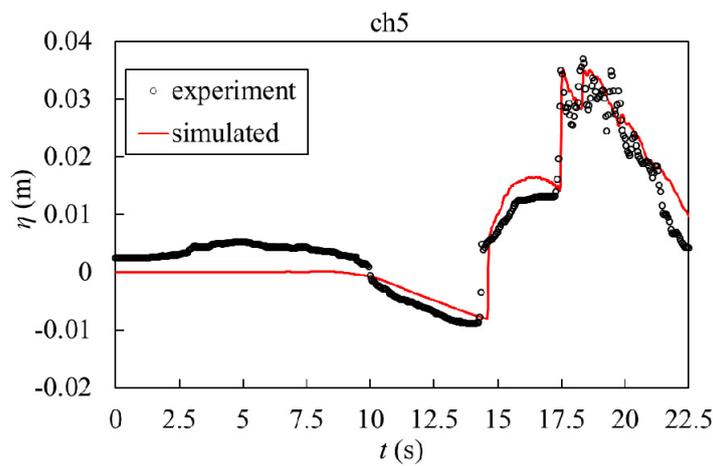

(a)

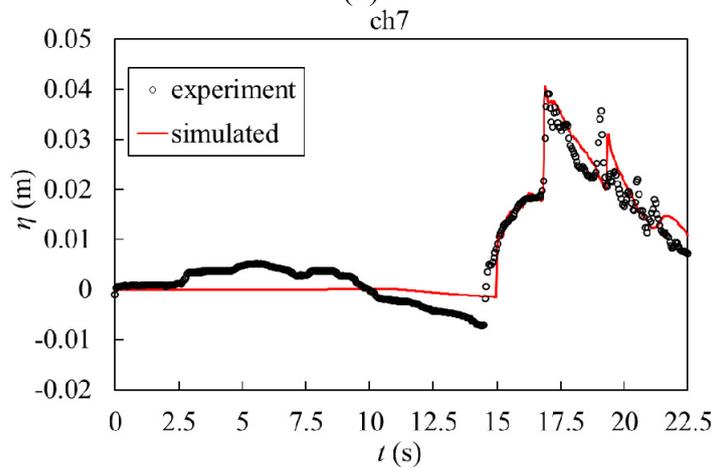

(b)

**Figure 5.** *Cont.*

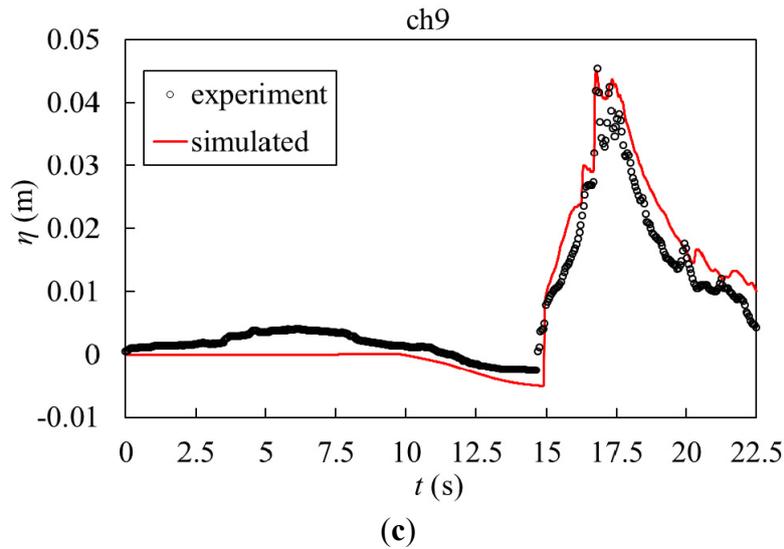

(c)

**Figure 5.** Comparison between computed and measured water elevations at different gauges (η represents water elevation). Numerical results are plotted in red lines and experimental measurements are plotted in black circles. (**a**–**c**) represent the water elevation varying with time at gauges ch5, ch7, and ch9, respectively.

*4.2. Case 2: One-Dimensional Dam Break Flow over Moveable Bed*

The experiment of a one-dimensional dam break flow over a movable bed conducted by Capart and Young [12] was used to test the capability of simulating the sediment erosion and bed changes using the present model. As shown in Figure 6, the experimental flume was 1.3 m in length, 0.2 m in width, and 0.7 m in depth and covered by sediment particles with a depth of 0.05–0.06 m. In this experiment, the sediment particles were light artificial spherical pearls covered with a shiny white coating. The uniform diameter of the particles was 6.1 mm with a density of 1.048 kg/m$^3$ and a porosity of 0.28. The initial water depth in the upstream was 0.1 m and the downstream was dry. At the beginning of the experiment, the gate was instantly removed to release the water, which generated flood propagation and caused erosion downstream.

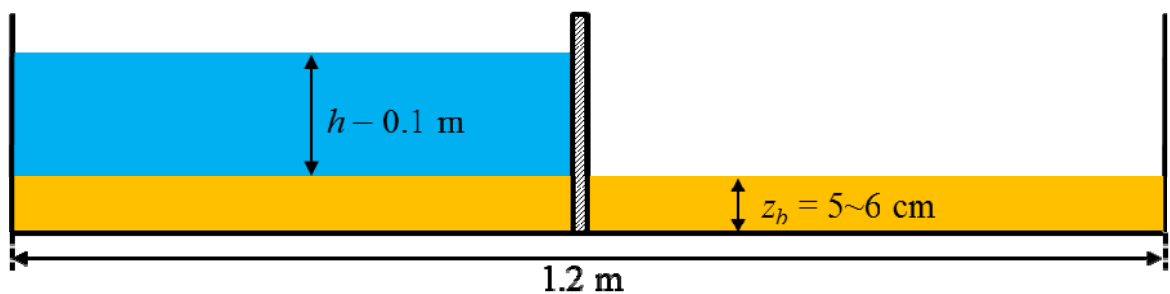

**Figure 6.** One-dimensional experiment of flood over movable bed.

The present model was used to simulate this process. In the simulation, the computational grid was 0.005 m and the Manning's coefficient was 0.025 sm$^{-1/3}$. The adaptation length $L_b$ and adaptation coefficient α were 0.2 m and 2.0, respectively. The settling velocity of sand used in the model was 7.6 cm/s based on the experimental data. The time step was adapted by Equation (37).

Figure 7 compares the numerical and experimental water surface and bed elevation at different time steps. As can be seen, the maximum erosion occurred downstream close to the gate, and the eroded area expanded downstream through flood propagation. A hydraulic jump was also observed near the gate of the dam. The erosion magnitudes and wavefront locations are well predicted by the present model. However, the location of the hydraulic jump is less accurately predicted. The main reason can be attributed to the present depth-averaged 2D model, which cannot precisely describe the three-dimensionality of the dam break flow over the movable bed. Overall, the simulated water depth and bed elevation agree well with the measured values. The relative errors of the bed elevation are less than 5.6%. Figure 8 shows the simulated sediment transport rate at different times. The maximum sediment concentration appeared at the front of the flood, where the high velocity can entrain more sediment to move with the water.

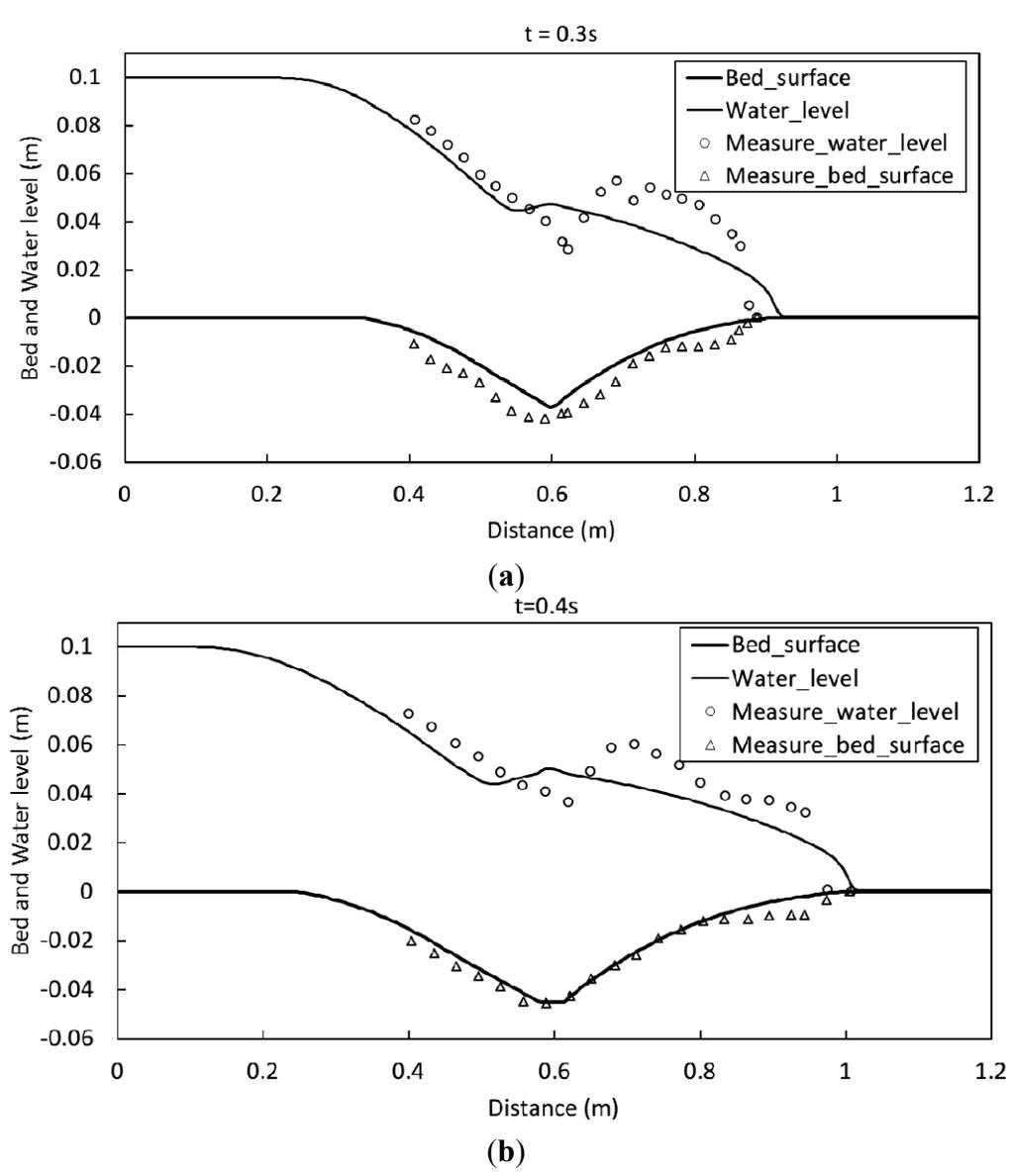

**Figure 7.** *Cont.*

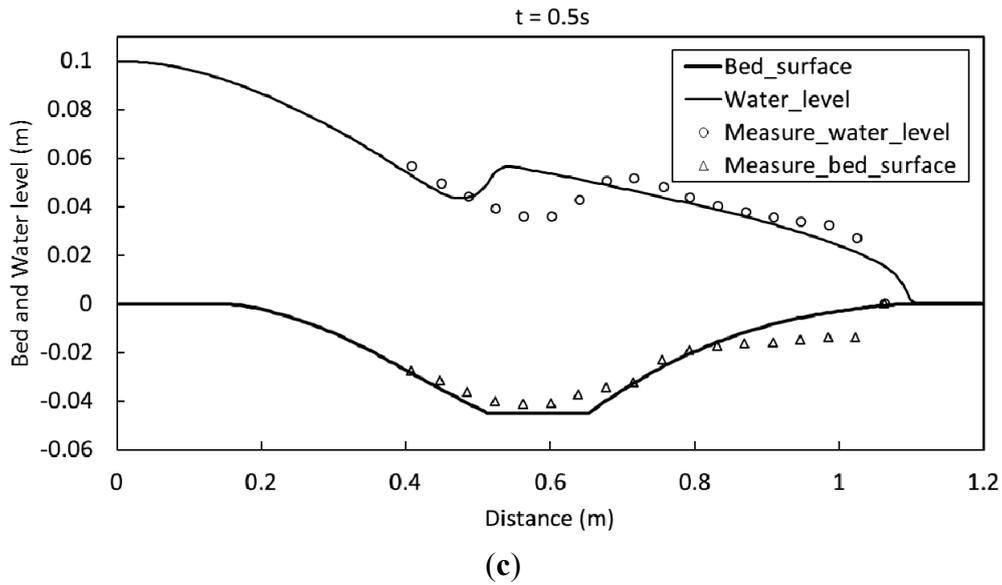

**Figure 7.** Comparison between simulated and measured water elevation and bed profiles. (**a**–**c**) compare the results at different times of 0.3 s, 0.4 s, and 0.5 s, respectively.

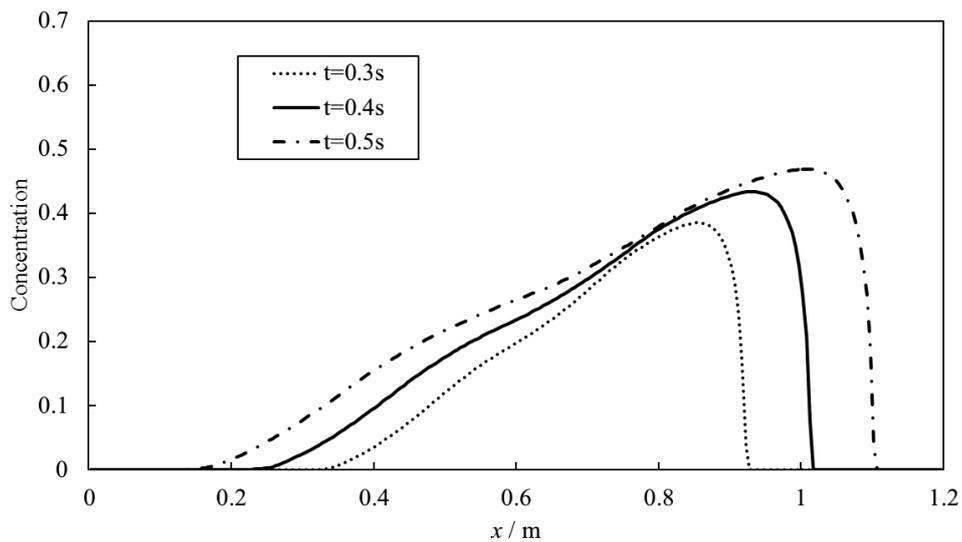

**Figure 8.** Simulated sediment concentration along the flood direction at different times.

## 5. Numerical Investigation of 2D Dam Breaching Processes with Overtopping Flow and Waves

The present model has been applied to simulate dam breaching experiments with and without waves in order to analyze the different features of sediment transport and erosion during different overtopping processes.

*5.1. Breaching Processes Due to Overtopping Flow*

The experiment considering flow overtopping without waves was carried out by HR Wallingford in the UK [31]. This non-wave case represents the overflow and erosion due to water level rises. The experiment was conducted in a flume with a dimension of 50 m in length and 10 m in width. The sandy dam, located upstream about 36 m towards the flume entrance, had a height of 0.5 m and an initial crest of 0.2 m in width and 0.02 m in depth at the top. The dam was made of non-cohesive sand with a median diameter of 0.25 mm. Both upstream and downstream slopes were 1:1.7. The initial water level was 0.5 m during the experiment, and the flow then overtopped the dam through the initial breach.

A rectangular grid of 1000 × 200 nodes was used in the simulation. The porosity of the dam was set to 0.36 based on the measurements, and the Manning's coefficient was 0.018. The initial water surface used in the simulation was 0.5 m. In the simulation, the adaptation length was 0.05 m, which was close to the grid space to avoid anti-dune formation in the numerical results [15]. The time step was determined by Equation (37). The inflow discharge of 0.07 m$^3$/s was specified at the flume entrance, while a free overall flow condition was applied at the end of the flume.

Figure 9 compares the measured and simulated breach flow discharges varying with time. The simulation results generally agreed well with the measured data, with a relative error of peak discharge of less than 8.2%, which shows that the model can provide a good prediction for such a complicated breaching process. Figure 10 shows the bed changes due to flow overtopping at the cross-section of the breach. The erosion at the initial stage was dominated by the vertical erosion and then changed to the lateral erosion, which means the breach became deeper at the beginning, and then became wider.

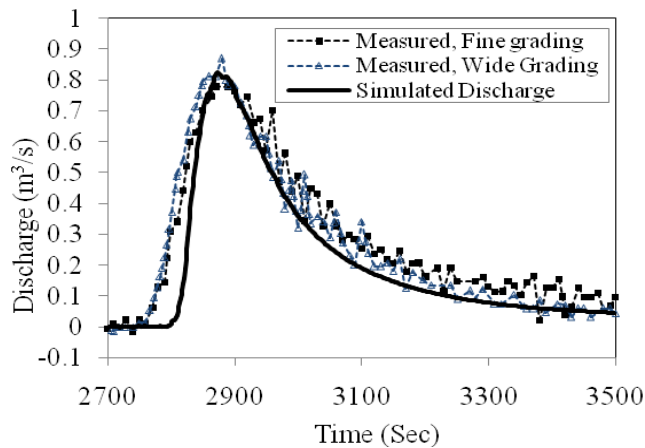

**Figure 9.** Measured and simulated breach flow discharges.

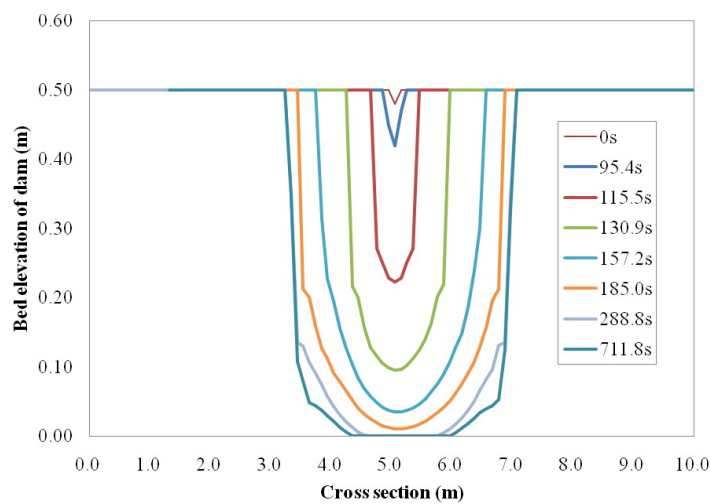

**Figure 10.** Simulated breach width and depth changes with no wave.

Figure 11 shows the simulated bed elevations at different time steps. Due to the rise of the water level, water mainly flowed through the initial breach. Gradually, the breach became wider and deeper and the sediment was deposited along two sides of the bank downstream.

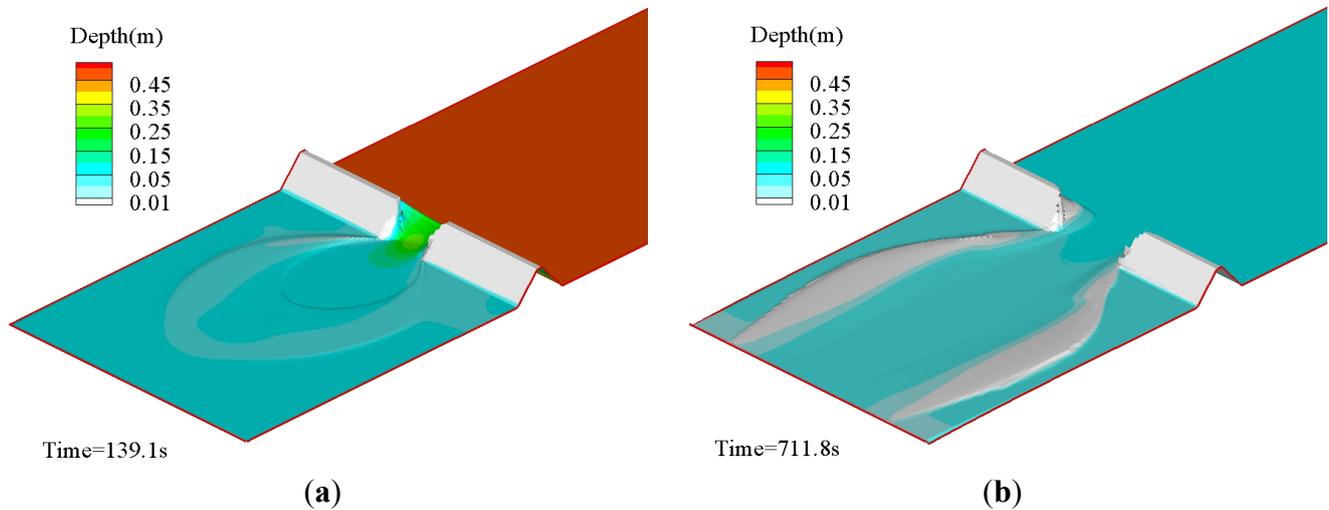

**Figure 11.** Bed changes and water depth with no wave at different times. The simulated water depths are plotted in the colored contours in the 3D topography view. The simulated changes of bed elevation and breach width and depth can be seen from (**a**) 139 s and (**b**) 712 s.

*5.2. Breaching Processes Due to Wave Overtopping*

To investigate the different behaviors of sediment transport and erosion by wave overtopping, a hypothetical numerical analysis of wave overtopping is carried out in this section for comparison. Unlike the water level steadily rising from the still water surface in the no-wave case above, this case assumes a continuous wave reaches the dam from the upstream. Other than this, the numerical analysis in this section keeps the same conditions as the previous experimental case. In the simulation, a regular wave with a wave height of 0.02 m and period of 1.5 s was applied at the upstream, while the initial water surface was 0.49 m. Therefore, the wave runup can overtop the dam. The computational mesh and other parameters used in the simulation were the same as the case of overtopping flow above, while the wave-adjusted boundary condition was used [32,33].

*5.3. Discussion of Results*

Figure 12 shows the simulated bed elevations at different time steps during the wave overtopping process. As can be seen, the flow overtopped the entire length of the sandy dune due to the wave attack and, correspondingly, erosion took place along the entire sandy dune in comparison to the breach erosion in Figure 11. Meanwhile, the erosion process was much faster than the case without waves and the initial breach at the top of the dam had little influence on the flow and erosion processes.

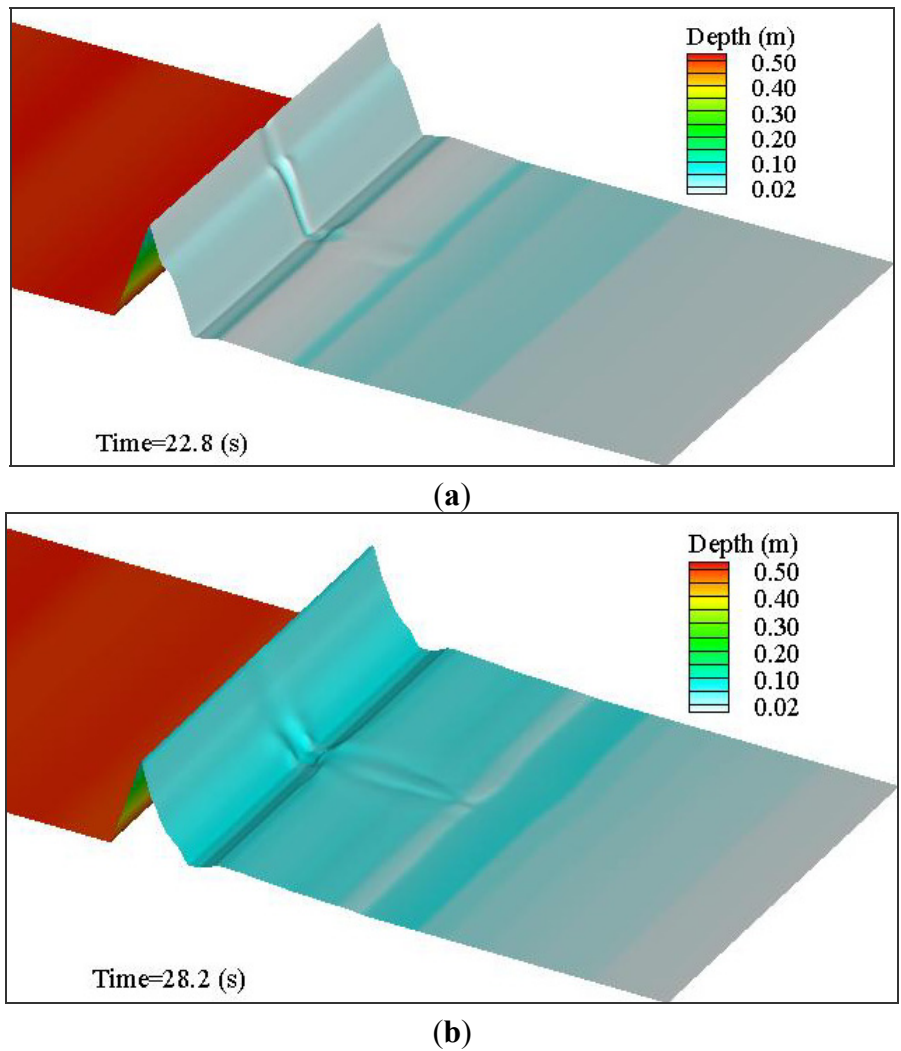

**Figure 12.** Bed changes and water depths with a wave at different times. The simulated water depths are plotted in the colored contours in the 3D topography view: (**a**) bed elevation and water depth at 22.8s, the sandy dam was eroded along the entire length and sediment was deposited in the downstream; (**b**) bed elevation and water depth at 28.2 s, the height of the dam was reduced quickly and erosion in the lower part of the dam downslope was increased.

To further see the different morphological changes, Figure 13 compares the bed changes at the middle cross-section along the flow direction during both flow and wave overtopping processes. As can be seen in Figure 13a, the downstream slope was eroded first and then the erosion expanded to the upper part during the flow overtopping. However, during the wave overtopping, as shown in Figure 13b, the erosion mainly occurred at the upper part at the initial stage because the wave ran up and overtopped the dune to cause a high flow velocity near the top of the dam. As the height of the dune reduced and the overtopping flow discharge increased, the entire dune was eroded fast as well. This process was significantly different in comparison to the one without the wave since the flow can only pass through the breach as shown in Figure 11.

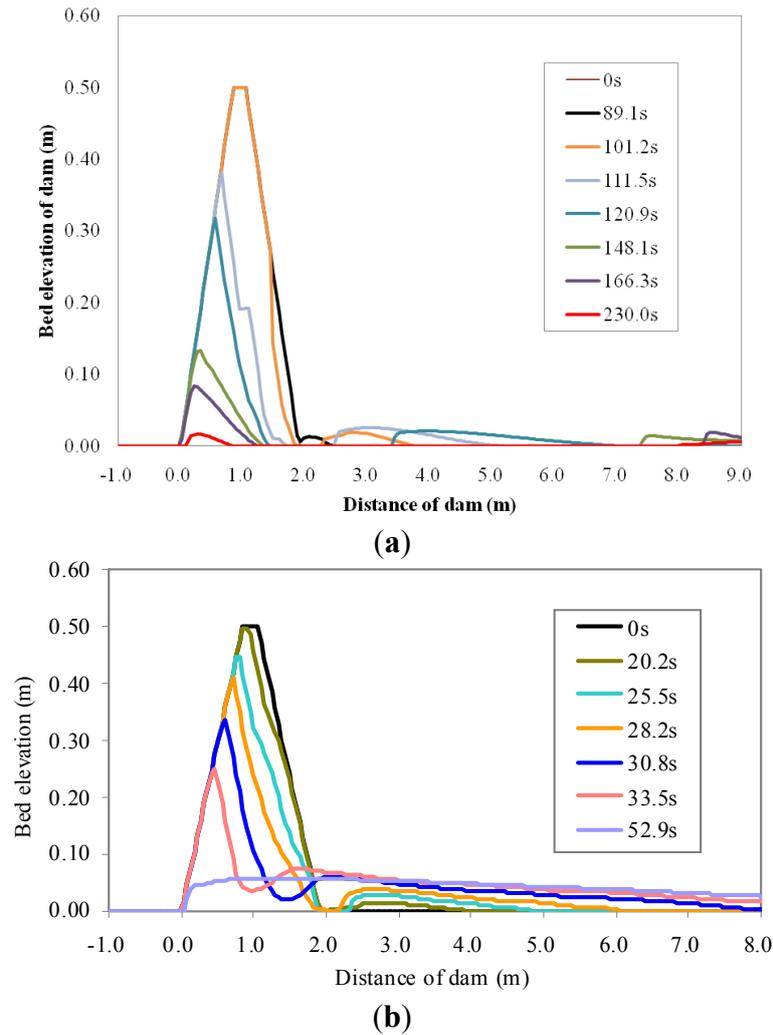

**Figure 13.** Comparison of bed changes along the flow direction at different times with and without waves. Numerical results of bed elevation along the central flow direction are plotted in the colored lines: (**a**) bed elevation changing with time in the case with no wave; (**b**) bed elevation changing with time considering overtopping waves.

## 6. Conclusions

A depth-averaged 2D coupled flow and sediment model is developed to investigate the earthen dam/levee breach processes due to overtopping flows with and without waves. The effects of sediment concentration and bed change on the flow are considered in the continuity and momentum equations, while the non-equilibrium transport of the total load is used for the sediment transport and bed changes. A specially designed avalanching algorithm is adopted in this model to consider different repose angles for the submerged materials and the dry materials above the water surface. The flow, sediment transport, and bed change equations are solved by the explicit algorithm based on the finite volume method. The model improves the Godunov-type central upwind scheme with nonnegative water depth reconstruction and a slope-limited linear reconstruction to efficiently handle the mixed-regime flows near the breach and the wetting-drying problem. To enhance the performance of the model, a varying time step is used that satisfies the CFL condition and ensures that the bed change at each time step is less than 10% of the local flow depth. The developed model has been tested using two sets of experimental data. The model predicts the water levels, bed changes, breach width, flow patterns near the breach, and the breach flow hydrograph generally well.

The numerical investigation of the breaching process due to overtopping flow with and without waves shows that waves have significant impacts on flow field, breaching width, depth, and the cross-section shape. In the case without waves, the flow through the breach eroded the downstream slope first. The erosion at the breach was dominated by the vertical erosion at the initial stage and then changed to the lateral erosion. In the wave overtopping case, the water overtopped the entire length of the dam due to the wave runup and caused the erosion along the entire length. Meanwhile, the erosion mainly occurred at the upper part at the initial stage. As the height of the dune reduced and flow discharge increased, erosion along the entire dune was faster than that in the no-waves case. However, it should be noted that the numerical results with waves are hypothetical and should be further verified by experiments or filed observations.

## Acknowledgments

This project is partially supported by the Natural Science Foundation of China (41376095, 41350110226), Scientific Research Foundation for the Returned Overseas Chinese Scholars (20101174), and Zhejiang University Ocean Sciences Seed Grant (2012HY012B). The authors wish to thank Weiming Wu for his help and discussions on sediment transport modeling over the years.

## Author Contributions

Zhiguo He had the original idea for the study, contributed to the mathematical model and drafted the manuscript. Peng Hu contributed to the mathematical model and numerical simulation, and helped with the analysis. Liang Zhao carried out the simulation and data analysis. Gangfeng Wu contributed to the numerical schemes, developed the numerical model, and contributed to numerical analysis and writing. Thomas Pähtz directed the simulations and provided geophysical expertise. All authors read and approved the final manuscript.

## Conflicts of Interest

The authors declare no conflict of interest.